\documentclass[conference]{IEEEtran}
\IEEEoverridecommandlockouts
\usepackage{cite}
\usepackage{amsmath,amssymb,amsfonts}
\usepackage{algorithmic}
\usepackage{graphicx}
\usepackage{multirow}
\usepackage[caption=false,font=normalsize,labelfont=sf,textfont=sf]{subfig}
\usepackage{textcomp}
\usepackage{xcolor}
\usepackage{soul}
\def \baselinestretch{0.92} 
\def\BibTeX{{\rm B\kern-.05em{\sc i\kern-.025em b}\kern-.08em
    T\kern-.1667em\lower.7ex\hbox{E}\kern-.125emX}}

\makeatletter
\newcommand{\linebreakand}{%
  \end{@IEEEauthorhalign}
  \hfill\mbox{}\par
  \mbox{}\hfill\begin{@IEEEauthorhalign}
}
\makeatother
    
\begin{document}

\title{2.5D Super-Resolution Approaches for X-ray Computed Tomography-based Inspection of Additively Manufactured Parts\\
\thanks{This manuscript has been authored by UT-Battelle, LLC, under contract DE-AC05-00OR22725 with the US Department of Energy (DOE). Research sponsored by the US Department of Energy, Office of Energy Efficiency and Renewable Energy, Advanced  Materials and Manufacturing Technologies Office (AMMTO), under contract DE-AC05-00OR22725 with UT-Battelle, LLC.
The US government retains and the publisher, by accepting the article for publication, acknowledges that the US government retains a nonexclusive, paid-up, irrevocable, worldwide license to publish or reproduce the published form of this manuscript, or allow others to do so, for US government purposes. DOE will provide public access to these results of federally sponsored research in accordance with the DOE Public Access Plan (http://energy.gov/downloads/doe-public-access-plan).}
}

\author{\IEEEauthorblockN{Haley Duba-Sullivan}
\IEEEauthorblockA{\textit{National Security Sciences Directorate} \\
\textit{Oak Ridge National Laboratory}\\
Oak Ridge, USA \\
sullivanhe@ornl.gov}
\and
\IEEEauthorblockN{Obaidullah Rahman}
\IEEEauthorblockA{\textit{Energy Science \& Technology Directorate} \\
\textit{Oak Ridge National Laboratory}\\
Oak Ridge, USA \\
rahmano@ornl.gov}
\linebreakand
\IEEEauthorblockN{Singanallur Venkatakrishnan}
\IEEEauthorblockA{\textit{Energy Science \& Technology Directorate} \\
\textit{Oak Ridge National Laboratory}\\
Oak Ridge, USA \\
venkatakrisv@ornl.gov}
\and
\IEEEauthorblockN{Amirkoushyar Ziabari}
\IEEEauthorblockA{\textit{Energy Science \& Technology Directorate} \\
\textit{Oak Ridge National Laboratory}\\
Oak Ridge, USA \\
ziabariak@ornl.gov}
}

\maketitle

\begin{abstract}X-ray computed tomography (XCT) is a key tool in non-destructive evaluation of additively manufactured (AM) parts, allowing for internal inspection and defect detection. Despite its widespread use, obtaining high-resolution CT scans can be extremely time consuming. This issue can be mitigated by performing scans at lower resolutions; however, reducing the resolution compromises spatial detail, limiting the accuracy of defect detection. 

Super-resolution algorithms offer a promising solution for overcoming resolution limitations in XCT reconstructions of AM parts, enabling more accurate detection of defects. While 2D super-resolution methods have demonstrated state-of-the-art performance on natural images, they tend to under-perform when directly applied to XCT slices. On the other hand, 3D super-resolution methods are computationally expensive, making them infeasible for large-scale applications. 

To address these challenges, we propose a 2.5D super-resolution approach tailored for XCT of AM parts. Our method enhances the resolution of individual slices by leveraging multi-slice information from neighboring 2D slices without the significant computational overhead of full 3D methods. Specifically, we use neighboring low-resolution slices to super-resolve the center slice, exploiting inter-slice spatial context while maintaining computational efficiency. This approach bridges the gap between 2D and 3D methods, offering a practical solution for high-throughput defect detection in AM parts.
\end{abstract}


\section{Introduction}

Additive manufacturing (AM) has revolutionized metal part production by enabling the construction of complex geometries that are difficult to achieve using traditional manufacturing methods. 
However, the quality of AM parts must be carefully evaluated, as defects such as pores, cracks, or inclusions can form during the printing process. 
X-ray computed tomography (XCT) is a essential tool for non-destructive evaluation of such parts, allowing for internal inspection and defect detection that ensures part reliability and performance.

Despite its usefulness, XCT struggles to achieve high-resolution scans of metal AM parts.
In particular, scanning thicker and more complex AM parts requires high-energy X-ray sources for better X-ray penetration of the material, which increases the spot size, thereby reducing resolution and introducing artifacts. 
Additionally, scanning large objects at high resolution is often impractical, as they may not fit within the field of view (FoV) and can require multiple stitched XCT scans, a process that is time-consuming, labor-intensive, and costly.
On the other hand, performing scans at lower resolutions -- either through geometric magnification in cone-beam CT or by binning the detector -- can speed up the process but results in a loss of spatial detail. 
In addition, binning is routinely used to enhance signal-to-noise (SNR) ratio during acquisition, thereby enhancing the reconstruction quality, which again will compromise the resolution.
These trade-offs limit the precision of defect detection, especially when hundreds of parts require rapid and cost-efficient characterization in industrial settings.

Supervised deep learning (DL)-based image-domain super-resolution methods have recently emerged as a promising solution to improve image resolution in XCT reconstructions~\cite{bashir2021comprehensive}. 
These techniques can enhance the resolution post-reconstruction. 
While 2D super-resolution methods achieve cutting-edge performance on benchmark natural RGB image datasets \cite{chen2024ntire}, directly applying them to individual slices of XCT volumes often leads to under-performance \cite{chatterjee2024beyond, chao2022generating}. 
In contrast, 3D methods can simultaneously enhance in-plane and through-plane resolution, but they are computationally expensive, limiting their applicability \cite{chen2024diffusion}.

To mitigate these limitations, we propose a 2.5D super-resolution approach tailored for XCT of AM parts. 
By exploiting multi-slice information from neighboring slices, our approach enhances the resolution of individual slices while avoiding the significant computational expense of full 3D methods. 
Specifically, we use six neighboring slices to super-resolve the center slice, allowing us to leverage inter-slice spatial context while maintaining the computational efficiency of 2D methods. 
This approach improves upon 2D methods while avoiding an increase in computational complexity as in 3D methods, providing a practical solution for high-throughput defect detection in metal AM parts.

\section{Related Work} \label{sec:related_work}
In this section, we will outline recent work on DL-based super-resolution methods, as well as previous work on 2.5D DL architectures. 

\subsection{Supervised DL-based Image-Domain Super-Resolution} \label{subsec:DL_SR}

% SRCNN and VDSR
DL methods have shown state-of-the-art performance for super-resolving natural images. 
Dong et al.~\cite{srcnn} proposed SRCNN as one of the first DL-based super-resolution methods.
SRCNN is a fully convolutional neural network that learns an end-to-end mapping between interpolated low- and high-resolution images using three convolutional layers. 

% ESPCN
While SRCNN was a significant step forward for super-resolution methods, the input images require bicubic upsampling of the low-resolution image first, which increases the computational complexity of the network. 
Shi et al.~\cite{shi_espcn} proposed an efficient sub-pixel convolutional neural network (ESPCN) which does the majority of the computation in the low-resolution space, significantly reducing the computational complexity compared to SRCNN. 
In this network, the input is the measured low-resolution image and upsampling filters are learned for each feature map and applied at the last layer. 

% SRResNet and EDSR
Residual learning, originally proposed in ResNet~\cite{resnet}, has been very successful in high-level computer vision problems like classification and detection. 
In 2017, Ledig et al. applied residual learning to super-resolution in SRResNet~\cite{Ledig_SRGAN_and_srresnet}. 
However, Lim et al.~\cite{lim_edsr} noted that modifications must be made to this architecture to optimize the network for super-resolution. 
They proposed an enhanced deep super-resolution network (EDSR) which removes unnecessary modules from the SRResNet structure. 
Namely, EDSR removes batch normalization layers which increases performance and decreases GPU memory usage. 

% SRGAN & ESRGAN
Ledig et al.~\cite{Ledig_SRGAN_and_srresnet} harnessed the promise of generative adversarial networks (GANs) in SRGAN. 
SRGAN uses SRResNet as the generator with a novel perceptual loss using high-level feature maps of a VGG network~\cite{simon_vgg} rather than the traditional MSE loss (which tends to create blurry images). 
Wang et al.~\cite{wang_esrgan} proposed an improvement on the original SRGAN called Enhanced SRGAN (ESRGAN). 
ESRGAN improves the original model in three specific ways -- the network structure, the discriminator, and the perceptual loss. 
For the network structure, the authors introduced the Residual-in-Residual Dense Block (RRDB) by combining multi-level residual networks and dense connections, remove Batch Normalization layers (as in~\cite{lim_edsr}), and use residual scaling and smaller initialization. 
For the discriminator, the authors used a relativistic average GAN~\cite{jolicoeur2018relativistic} which helps the generator recover more realistic texture detail. 
Finally, the authors improved the perceptual loss by using the VGG features before activation instead of after activation. 
ESRGAN is shown to consistently achieve better perceptual quality than previous super-resolution methods, including SRGAN.

\begin{figure}
\centering
\includegraphics[width=0.43\textwidth]{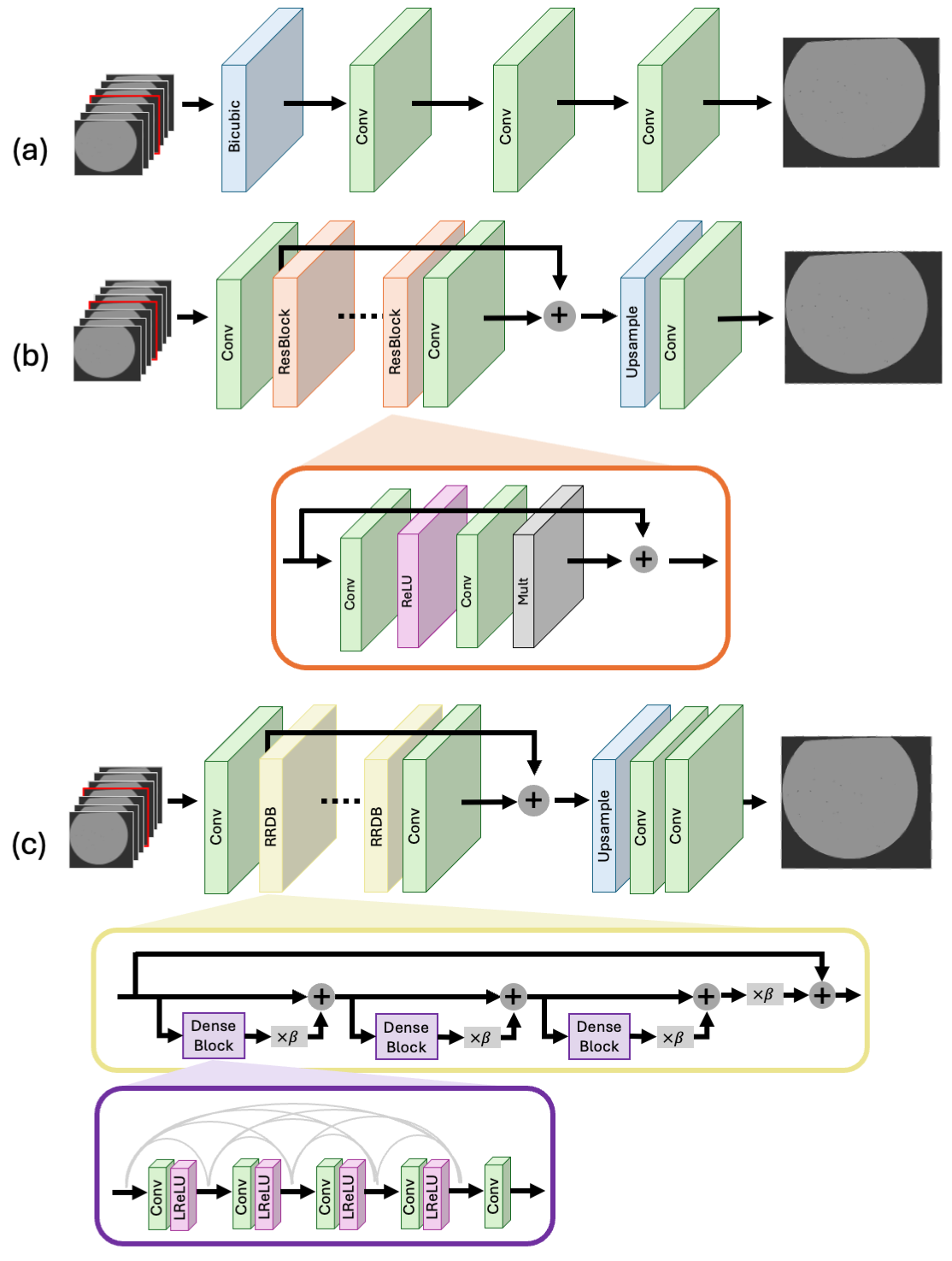}
\caption{2.5D Architectures for (a) SRCNN~\cite{srcnn}, (b) EDSR~\cite{lim_edsr}, and (c) ESRGAN generator~\cite{wang_esrgan}. We use the standard 2D discriminator for ESRGAN, as presented in \cite{wang_esrgan}.}
\label{fig:2.5D Architectures}
\end{figure}

\section{Experimental Results}\label{sec:results}
In this section, we present results comparing 2D, 2.5D, and 3D super-resolution using three different networks -- SRCNN~\cite{srcnn}, EDSR~\cite{lim_edsr}, and ESRGAN~\cite{wang_esrgan}.

\subsection{Datasets}\label{subsec:data}

\begin{table}
\caption{Synthetic and real dataset generation:  Materials, number of views, detector binning and reconstruction technique used to create the reconstruction volumes for training and testing.}
\label{tab:data_specs}
\begin{center}
\begin{tabular}{| c || c c c c c|}
\hline \rule{0pt}{1.0\normalbaselineskip}
Dataset & Input/Target & Material & Num. views & Binning & Recon \\
\hline \rule{0pt}{1.0\normalbaselineskip}
\multirow{2}{*}{Synthetic} & Input & Al. & 1066 & 4$\times$ & FDK \\
& Target & Al. & 2132 & -- & MBIR \\
\hline \rule{0pt}{1.0\normalbaselineskip}
\multirow{2}{*}{Real} & Input & St. & 147 & 4$\times$ & FDK \\
& Target & St. & 1066 & -- & MBIR \\
\hline
\end{tabular}
\end{center}
\end{table}

% 2D and 3D limitations
While these 2D DL-based super-resolution methods have shown state-of-the-art performance for super-resolving natural images, they disregard multi-slice information when applied to XCT reconstructions of AM parts, which degrades performance. 
Extending these to 3D super-resolution methods requires 3D convolutional kernels, which increases the computational complexity and the required compute resources, limiting applicability.
The limitations of 2D and 3D methods motivate the development of 2.5D DL architectures which exploit multi-slice information without significantly increasing computational complexity.

\subsection{2.5D DL Architectures} \label{subsec:2.5D_DL}

Previous work has demonstrated the ability of 2.5D DL architectures to produce image fidelity comparable to 3D DL methods while offering computational complexity similar to 2D DL methods. 
2.5D DL has been used for CT image reconstruction~\cite{2.5DMBIR, ziabari2023enabling, ziabari2022simurgh, rahman2023mbir,majee2021multi}, CT image denoising~\cite{2.5Ddenoising, rahman2023deep}, and volumetric medical image segmentation~\cite{2.5Dsegmentation}.
Despite the success of 2.5D DL in balancing image fidelity with computational complexity, to the best of our knowledge, it has not been used for super-resolving XCT reconstructions of AM parts. 

\section{2.5D Super-Resolution}\label{sec:2.5D SR}
In this section, we introduce a 2.5D architecture for DL-based super-resolution methods. 
Rather than proposing a specific 2.5D super-resolution method, we propose a simple architecture alteration that can be applied to any existing super-resolution network.
This general 2.5D architecture allows incorporation of multi-slice information into any DL-based super-resolution method, which enables easy adaptation of cutting-edge super-resolution research to XCT of AM parts.

By leveraging multi-slice information from neighboring slices, the 2.5D architecture enhances the resolution of individual slices without the significant computational overhead associated with full 3D methods.
Namely, we alter a 2D super-resolution network to take the center slice and 6 neighboring slices (7 contiguous slices) as input and output the super-resolved center slice. 
The full super-resolved volume is then generated using a sliding window of 7 slices in one dimension. 
Note that any odd number can be used for the number of input slices but 7 was selected because it gave the best results. 

We select three DL-based super-resolution methods to demonstrate our 2.5D architecture.
Namely, we use SRCNN~\cite{srcnn}, EDSR~\cite{lim_edsr}, and ESRGAN~\cite{wang_esrgan}, which are discussed in Section~\ref{subsec:DL_SR}. 
% Although SRCNN is not a state-of-the-art method, it is a lightweight CNN which can be important in some applications. 
% EDSR and ESRGAN are commonly used super-resolution methods \cite{furat2022super, jackson2022deep, karamov2023super}.
Figure~\ref{fig:2.5D Architectures} shows the modified architectures for each of these networks. 
Note that the 2.5D ESRGAN architecture includes a 2.5D generator and a 2D discriminator. 

We present results on a synthetic and real dataset.
Table~\ref{tab:data_specs} outlines the generation parameters for the input (low-resolution) and target (high-resolution) volumes for both datasets, including the material, number of views, detector binning factor, and reconstruction technique. 

Our synthetic dataset consists of paired low- and high-resolution XCT volumes generated using Computer-Aided Design (CAD) models of two aluminum AM parts. 
The simulated detector size was set to $1456  \times 1840$ pixels, with each pixel measuring $0.127 \times 0.127$ mm$^2$, matching a standard detector being used in commercial industrial XCT systems (e.g. Zeiss Metrotom). 
We used python's spekpy package (\cite{spekpy1, spekpy2}), to simulate the XCT spectrum with a peak voltage of 180kV and a 2mm Al filter to reduce the beam hardening effect.
% Distances (between the source and iso-center, and source and detector) were chosen so that the high-resolution reconstructed volume has a voxel size of 17.28 $\mu$m.  
The high-resolution reconstructed volume has a voxel size of 17.28 $\mu$m.
After initial simulations, the detector was binned by $4\times$, resulting in reconstruction volumes with voxel size of 69.12 $\mu$m. 
The low-resolution volumes were generated with a scan of 1,066 views using FDK (Feldkamp, Davis and Kress)~\cite{fdk} with no beam hardening correction.
The high-resolution volumes were generated with a full scan of 2,132 views using MBIR (model-based iterative reconstruction)~\cite{yu2010fast, rahman2020direct} with beam hardening correction~\cite{rahman2023neural}. 
MBIR usually has desirable noise texture compared to FDK. 
The job of the networks, apart from super-resolving low-resolution images, is to remove the beam-hardening effect and match MBIR's texture which will enhance defect contrast/detectability.

Our real dataset consists of paired low- and high- resolution volumes generated from XCT scans of two steel AM parts. 
The detector size was $1456 \times 1840$ pixels, with each pixel measuring 0.127 × 0.127 mm$^2$. 
% Distances were chosen so that the high-resolution reconstructed volume has a voxel size of 17.28 $\mu$m.
The high-resolution reconstructed volume has a voxel size of 17.28 $\mu$m.
The low-resolution sinogram was generated by binning the detector (projection) data by $4\times$, which results in $4\times$ larger detector pixel size and in turn reconstruction volumes with voxel sizes of 68 $\mu$m.
The low-resolution volume was generated with a sparse scan of 147 views using FDK~\cite{fdk} with no beam hardening correction.
The high-resolution volume was generated with a scan of 1,066 views using MBIR with beam hardening correction \cite{rahman2023neural}.
Note that since steel is a dense material, it will result in more beam hardening, and therefore a more challenging dataset (especially when the number of views are significantly reduced).

\subsection{Implementation Details}

\begin{table*}[ht]
    \centering
    \caption{Comprehensive comparison of the number of trainable parameters and maximum memory allocated for 2D, 2.5D, and 3D architectures. Extending to a 3D architecture has a significant impact on the parameters and memory, while extending to a 2.5D architecture has a very small impact.}
    \label{tab:comprehensive_comparison}
    \begin{tabular}{|c|l||r|r|r||r|r|}
        \hline
        \rule{0pt}{1.1\normalbaselineskip}
        \multirow{2}{*}{Method} & \multirow{2}{*}{Metric} & \multicolumn{3}{c||}{Values} & \multicolumn{2}{c|}{Percent Increase Over 2D} \\
        \cline{3-7}\rule{0pt}{1.1\normalbaselineskip}
        & & \multicolumn{1}{c|}{2D} & \multicolumn{1}{c|}{2.5D} & \multicolumn{1}{c||}{3D} & \multicolumn{1}{c|}{2.5D} & \multicolumn{1}{c|}{3D} \\
        \hline
        \rule{0pt}{1.1\normalbaselineskip}
        \multirow{3}{*}{SRCNN} & Number of Parameters & 57,281 & 88,385 & 306,753 & 54.30\% & 435.52\% \\
        & Max. Memory Allocated (GB) & 0.1332 & 0.1367 & 16.6721 & 2.63\% & 12,416.59\% \\
        \hline
        \rule{0pt}{1.1\normalbaselineskip}
        \multirow{3}{*}{EDSR} & Number of Parameters & 1,515,265 & 1,518,721 & 3,655,169 & 0.23\% & 141.22\% \\
        & Max. Memory Allocated (GB) & 0.2007 & 0.2009 & 14.7148 & 0.10\% & 7,231.74\% \\
        \hline
        \rule{0pt}{1.1\normalbaselineskip}
        \multirow{3}{*}{ESRGAN} & Number of Parameters & 31,193,930 & 31,197,386 & 124,151,050 & 0.01\% & 298.00\% \\
        & Max. Memory Allocated (GB) & 6.2007 & 6.2010 & 75.8162 & 0.005\% & 1,122.70\% \\
        \hline
    \end{tabular}
\end{table*}

We used two different reconstructions for training and testing. 
Training patches were extracted with an in-plane stride of $64$ voxels such that the 2D high-resolution training patches were $1 \times 128 \times 128$ voxels, the 2.5D high-resolution training patches were $7 \times 128 \times 128$ voxels, and the 3D high-resolution training patches were $128 \times 128 \times 128$ voxels.

We use the standard architectures proposed by the authors for each network.
Namely, SRCNN has 3 convolutional layers, EDSR has 16 residual blocks and each convolutional layer has 64 feature maps, and  ESRGAN has 32 grow channels and 23 RRDB blocks with 64 feature maps.
For 3D versions of each network, we replace all 2D convolutional layers with 3D convolutional layers. 

Table~\ref{tab:comprehensive_comparison} compares the number of trainable parameters and maximum allocated memory for 2D, 2.5D, and 3D architectures. 
Extending to a 3D architecture from 2D increases the number of trainable parameters by more than 140\% and the memory allocated by over 1000\%.
On the other hand, extending to a 2.5D architecture increases the number of trainable parameters by less than 55\% and the memory allocated by less than 3\%.
Note that the increase in number of parameters from 2D (1 input channel) to 2.5D (7 input channels) is a constant with respect to the patch size.
Since each of these networks begins with a convolutional layer, this constant increase can be computed as $6mnk$, where $m\times n$ is the size of the kernel in the first convolutional layer and $k$ is the number of feature maps in the first convolutional layer.

\subsection{Results on Synthetic Dataset}

\begin{figure}
\centering
\includegraphics[width=0.48\textwidth]{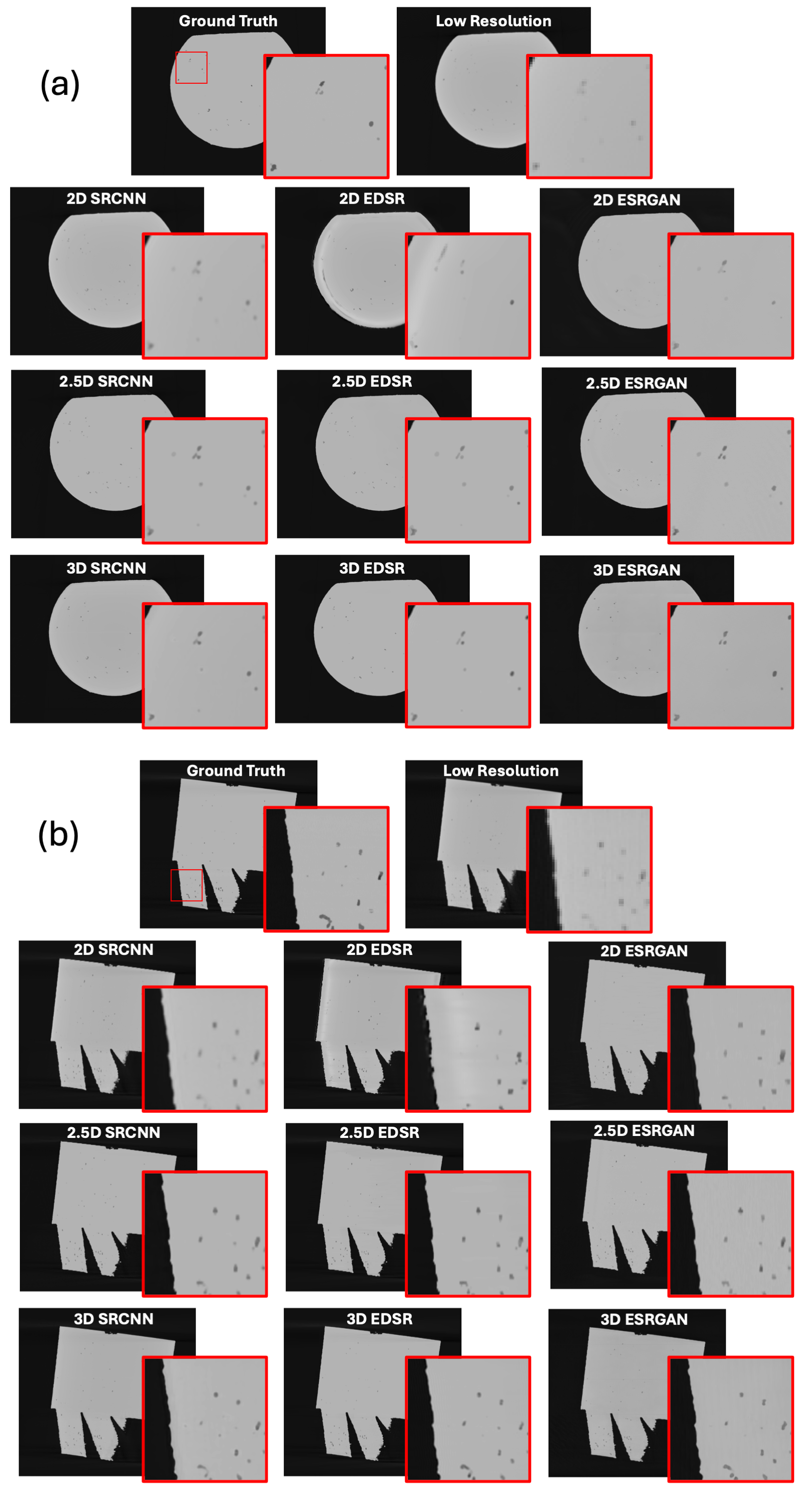}
\caption{Comparison of (a) XY- and (b) XZ-slices of super-resolved synthetic XCT volume using 2D, 2.5D, and 3D architectures. The 3D architectures produces the highest quality reconstructions at the cost of a significant increase in memory. The 2.5D architectures produce higher-quality reconstructions that better match the ground truth compared to the 2D architectures.}
\label{fig:synth}
\end{figure}

Figure~\ref{fig:synth} compares XY- and XZ-slices (YZ-slices excluded as they are similar to XZ-slices.) (axial and vertical cross-section slices respectively) of the super-resolved synthetic XCT test volume using 2D, 2.5D, and 3D architectures of SRCNN, EDSR, and ESRGAN. 
Visually, the 3D architectures generate the most accurate reconstructions. 
However, the 2.5D architectures perform similarly to 3D without significantly increasing the computational complexity of the method, and produce higher-quality reconstructions than the 2D architectures.
The visual differences between the super-resolved volumes generated with 2D and 2.5D architectures are more prominent in the XZ-slice (Figure~\ref{fig:synth}(b)) since the 2D methods do not learn any inter-slice spatial information in the Z dimension.

\begin{figure}
    \centering
    \includegraphics[width = 0.48 \textwidth]{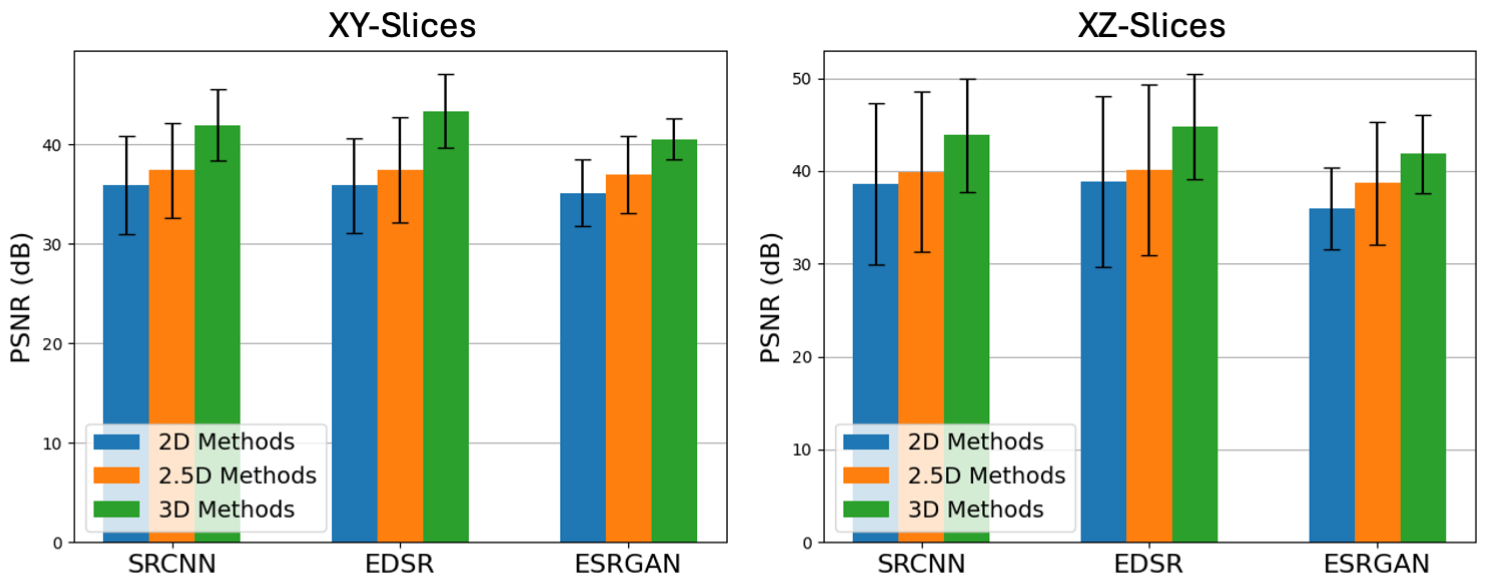}
    \caption{Mean and standard deviation of PSNR over XY- and XZ-slices of the super-resolved synthetic XCT volume using 2D, 2.5D, and 3D architectures. The 3D architectures attain the highest mean PSNR, but they also significantly increase the required memory. The 2.5D architectures attain a higher mean PSNR than the 2D architectures with a negligible increase in the required memory.}
    \label{fig:psnr}
\vspace{.01cm}
\end{figure}

Figure~\ref{fig:psnr} compares the mean and standard deviation of the PSNR over XY- and XZ-slices of the super-resolved volumes using 2D, 2.5D, and 3D architectures. 
As expected, the 3D architectures attain the highest mean PSNR, but this is at the cost of a significant increase in required memory.
Meanwhile, the 2.5D architectures attain a higher mean PSNR than the 2D architectures with a negligible increase in required memory. 
This directly supports our conclusion that 2.5D super-resolution produces better quality images than 2D super-resolution without significantly increasing complexity. 

\begin{figure*}
\centering
\includegraphics[width=0.95\textwidth]{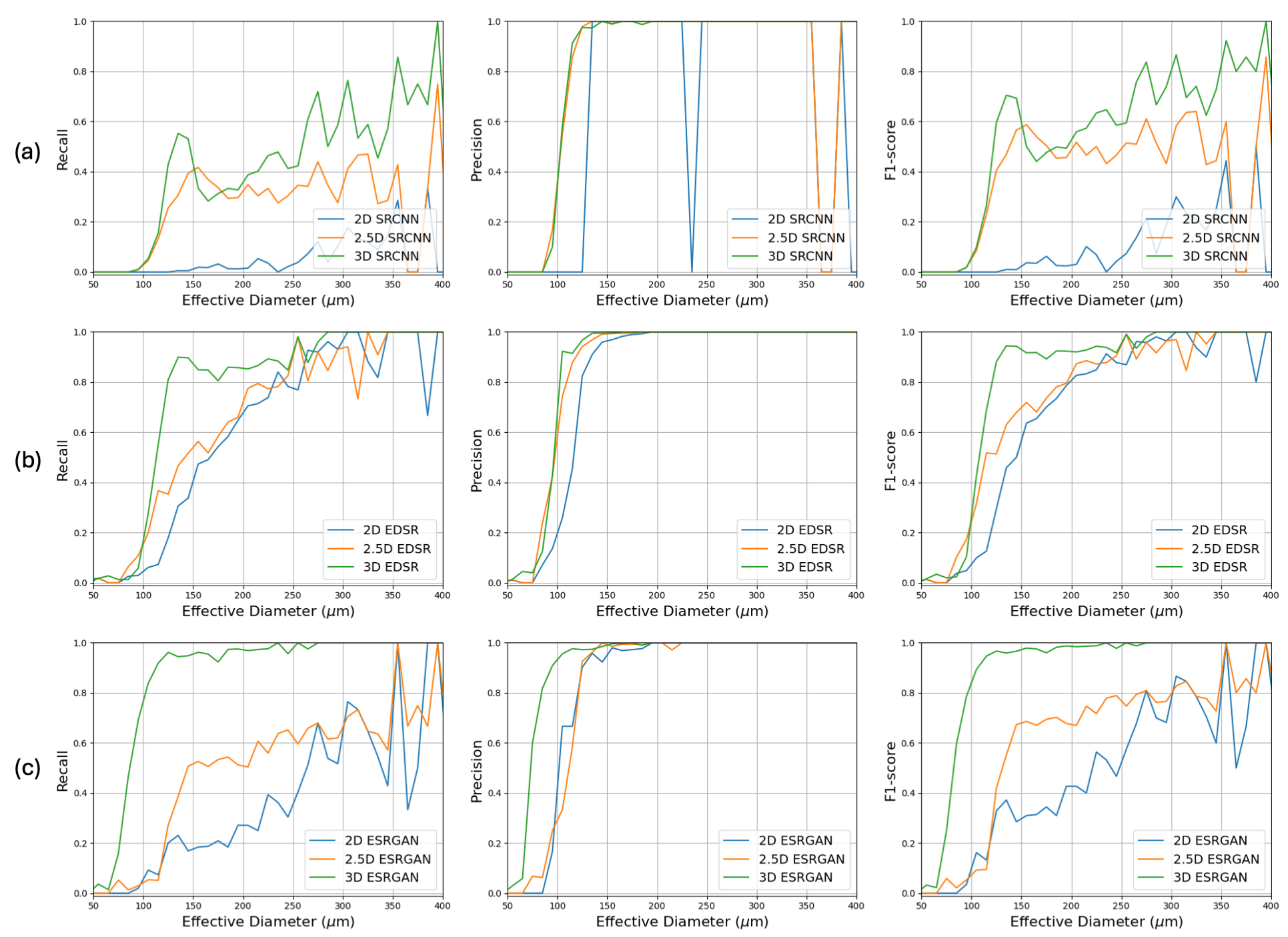}
\caption{Recall, precision, and F1 score curves for super-resolved volumes using 2D, 2.5D, and 3D architectures of (a) SRCNN, (b) EDSR, and (c) ESRGAN. The 3D architectures attain the highest recall, precision, and F1 score; however, they also increase the required memory by more than $1000\%$. The 2.5D architectures attain a higher recall, precision, and F1 score than the 2D architectures with a negligible increase in the required memory.}
\vspace{-0.3cm}
\label{fig:recall}
\end{figure*}

While successful super-resolution improves image quality, recall that our main motivation is to improve the ability to accurately detect defects from the super-resolved XCT reconstruction.
To quantify this, we calculate the recall, precision, and F1 score associated with defect detection from each of the super-resolved XCT reconstructions.
Figure~\ref{fig:recall} compares each of these values as a function of the effective defect diameter for the super-resolved volumes using 2D, 2.5D, and 3D architectures of SRCNN, EDSR, and ESRGAN.
Note that the precision and recall of 2D SRCNN drop to zero for an effective diameter of 240 $\mu$m, since all defects of that size went undetected. This is consistent with the overall poor detection performance of 2D SRCNN compared to the other methods.
Overall, the 3D architectures generally attain higher recall, precision, and F1 score than 2D and 2.5D (most noticeably for small defects), with a more significant impact on recall than precision. 
The 2.5D architectures consistently attain higher recall, precision, and F1 score than the 2D architectures, also with a more significant impact on recall than precision. 
This supports our conclusion that 2.5D super-resolution enables more accurate defect detection than 2D super-resolution.

\subsection{Preliminary Results on Real Dataset}
We also present preliminary results comparing the performance of 2D, 2.5D, and 3D ESRGAN on a real dataset. 
Real XCT reconstructions pose a much more difficult super-resolution problem due to the complex degradations present, including noise, blur, beam hardening, and streaking. 
In preliminary experiments, we have empirically observed that SRCNN and EDSR struggle to successfully super-resolve the real XCT reconstructions. 
For this reason, we only include preliminary ESRGAN results in this paper. 

\begin{figure}
\centering
\includegraphics[width = 0.48\textwidth]{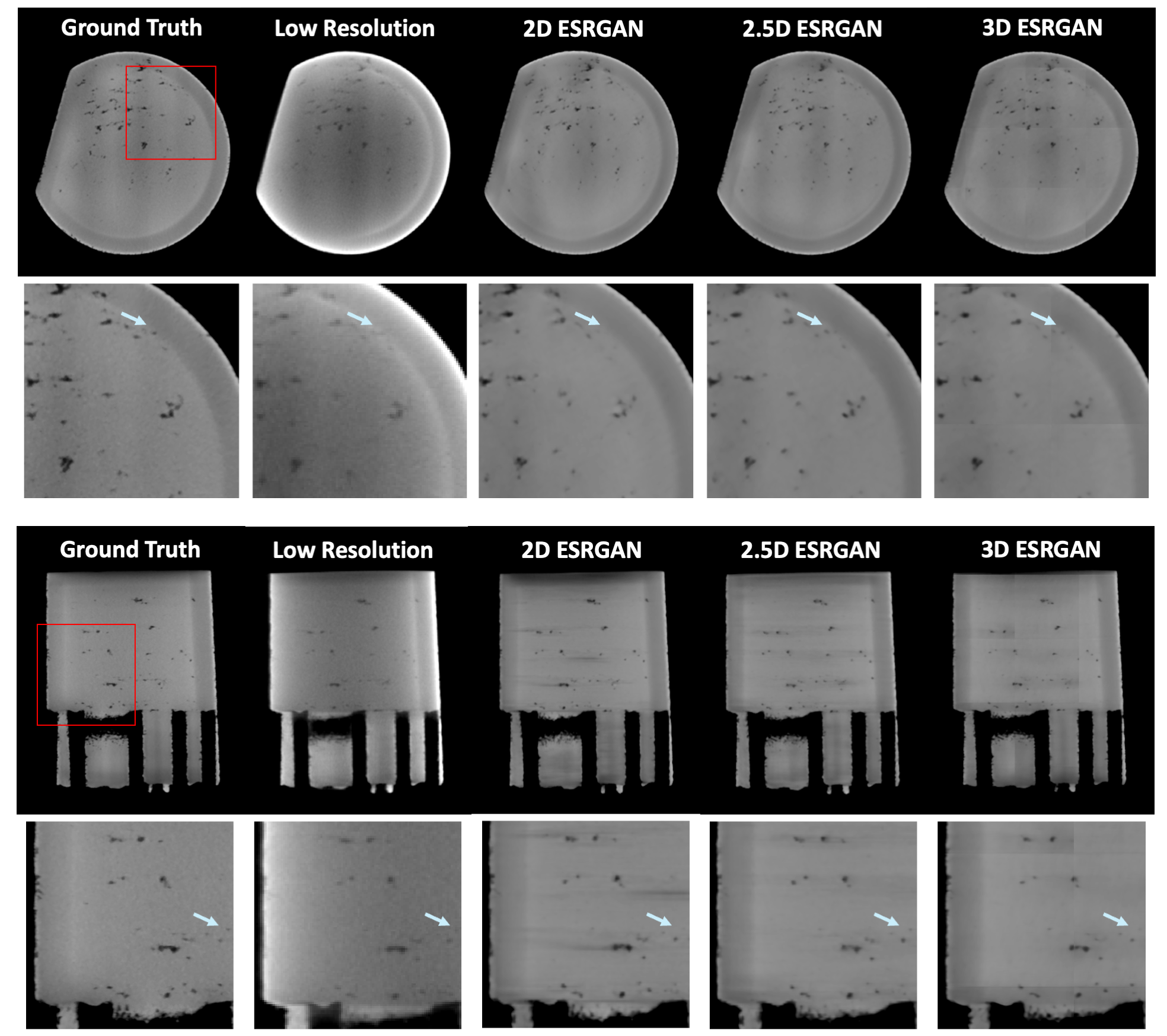}
\caption{Comparison of XY-slices (top) and XZ-slices (bottom) from preliminary results on super-resolved real XCT test volume using 2D, 2.5D, and 3D ESRGAN. Visually, 2.5D ESRGAN resolves small defects better than 2D ESRGAN and is on par with 3D ESRGAN.}
\vspace{-0.5cm}
\label{fig:real}
\end{figure}

Figure~\ref{fig:real} compares XY- and XZ-slices of the super-resolved real XCT test volume using 2D, 2.5D, and 3D ESRGAN. 
2.5D ESRGAN super-resolves small defects slightly better than 2D ESRGAN, as denoted by the arrows.
Additionally, 2.5D ESRGAN produces fewer artifacts in the Z dimension than 2D ESRGAN, as seen in Figure~\ref{fig:real}(b). 
3D ESRGAN has the fewest overall artifacts in the Z dimension, but visually performs similarly to 2.5D ESRGAN in super-resolving small defects.
Note that memory constraints prohibit super-resolving the entire 3D volume at once; visible grid artifacts in 3D ESRGAN arise as a result of stitching smaller volumes together to assemble the entire 3D volume.
In future work, we plan to perform more extensive testing on real data.

\section{Conclusion}

In this paper, we used 2.5D super-resolution to enhance the resolution of individual slices of XCT reconstructions of AM parts.
Our 2.5D super-resolution method leverages multi-slice information from neighboring slices without the significant computational overhead associated with full 3D methods. 
This simple architecture adjustment allows for easy adaptation of existing state-of-the-art 2D DL-based super-resolution methods to XCT volumes, offering boosts in image quality and defect detection without a significant increase in computational complexity.

In a set of experiments on a synthetic dataset, 2.5D super-resolution improved the image quality and the ability to detect defects when compared to 2D super-resolution while incurring less than a $3\%$ increase in required memory. 
In comparison, 3D methods required over $1000\%$ more memory than 2D methods.
While 2D super-resolution methods are able to attain good image quality as quantified by PSNR, they perform significantly worse than 2.5D and 3D super-resolution methods on task-specific metrics as quantified by recall, precision, and F1 score. 
Additionally, preliminary results on real data show that 2.5D outperforms 2D super-resolution and performs similarly to 3D super-resolution. 
Future work will include more extensive testing on real data and application of the proposed 2.5D architecture to more recent super-resolution methods, such as diffusion- and transformer-based methods.

\section*{Acknowledgments}
We would like to thank Gregery Buzzard and Charles Bouman from Purdue University for their support and constructive discussions throughout this project. 

\bibliographystyle{IEEEtran}
\bibliography{IEEEabrv,main}

\end{document}